\documentclass[proceedings]{JHEP}
\usepackage{epsfig}
\newcommand{\be}{\begin{eqnarray}}
\newcommand{\ee}{\end{eqnarray}}
\conference{Third Latin American Symposium on High Energy Physics}
\title{Chiral Ward identities and pion propagation at finite
temperature in the linear sigma model}
\author{Alejandro Ayala\thanks{In collaboration with S. Sahu and
                               M. Napsuciale}\\ 
        Instituto de Ciencias Nucleares, 
        Universidad Nacional Aut\'onoma de M\'exico,\\
        Apartado Postal 70-543, M\'exico Distrito
        Federal 04510, M\'exico.\\
        E-mail: ayala@nuclecu.unam.mx}  

\abstract{Working within the linear sigma model at finite temperature,
we construct effective one-loop vertices and propagators satisfying
the chiral Ward identities, for small pion momentum and mass
compared to the sigma mass. We obtain an expansion for the pion
dispersion relation up to second order in the parameter
$m_\pi^2/4\pi^2f_\pi^2$. At leading order, this expansion reproduces
the dispersion curve obtained in chiral perturbation theory. We also
study pion damping and compute the mean free path for pions traveling 
in the medium before forming a sigma
resonance. These results are also in good agreement with chiral perturbation
theory when using a value for the sigma mass of $600$ MeV.}


\begin{document}

\section{Introduction}

In recent years, a considerable effort has been devoted to the
production and detection of the so called quark-gluon plasma 
--the state of matter where the basic QCD degrees of freedom are not
confined to a single hadron-- in relativistic heavy-ion
collisions. The disentanglement of the signals to discern between
different outcomes of the collision, requires the understanding of the
evolution of the hadronic state from production to freeze out. Since
pions are the most copiously produced particles in this kind of
collisions, a proper understanding of their propagation properties
within a dense a hot environment represents a key ingredient for the 
understanding of the system's physical properties as a whole.

In general, the hadronic degrees of freedom are accounted for by means of 
effective chiral theories whose basic ingredient is the fact that pions are
the Goldston bosons coming from the spontaneous breakdown of chiral
symmetry. Chiral perturbation theory (ChPT) is one of such theories
that has been employed to show a number of properties
exhibited by pions at finite temperature, at leading~\cite{Gasser} and 
next to leading~\cite{{Schenk},{Toublan}} orders. A striking result 
obtained from the second order calculations is that the shift in the 
temperature dependence of the pion mass is opposite in sign and about
three times larger in magnitude than the first order shift, already at
temperatures of order of $150$ MeV. This result might signal either
the breakdown of the perturbative scheme at such temperatures or the
need to compute beyond next to leading order in ChPT, given the large
relative corrections found. 

Nevertheless, the simplest realization of chiral symmetry is still 
provided by the much studied linear sigma model~\cite{Schwinger}. 
This is a renormalizable field theory at zero~\cite{Lee} and 
(consequently) at finite temperature~\cite{Mohan}. The
model has been the subject of a renewed interest in view of recent   
theoretical results~\cite{Tornqvist} and analyses of data~\cite{Svec} 
that seem to confirm the existence of a broad scalar resonance, with a mass 
in the vicinity of 600 MeV, that can be identified with the $\sigma$-meson.
If confirmed, this last feature could place the linear sigma model
above the status of a toy model to study QCD below the deconfinement
scale. It thus becomes important to check that the results obtained by
means of other theories such as ChPT can be reproduced by the linear sigma
model when working in a given kinematical regime. 

Here we show that the linear sigma model reproduces the leading order
modification to the pion mass in a thermal pion medium obtained in ChPT at
leading order, when use is made of a systematic expansion in 
the ratio $m_\pi^2/m_\sigma^2$ at zeroth order --with $m_\sigma$, 
$m_\pi$ the vacuum sigma and pion masses, respectively-- and when
treating the momentum and temperature as small quantities of order $m_\pi$.
In this regime, the effective expansion parameter becomes 
$m_\pi^2/4\pi^2f_\pi^2$, with $f_\pi$ the pion decay constant. 
Such parameter allows for a controlled loop expansion of the pion
self-energy from where we can study both, the modification to the pion
mass at next to leading order and pion damping in the elastic and
inelastic channels. This is accomplished by constructing 
one-loop effective vertices and propagators satisfying the chiral Ward
identities and thus respecting chiral symmetry. Details of the
calculation as well as a more extended discussion on its implications
can be found in Refs.~\cite{Ayala}.

\section{Effective vertices and chiral Ward identities}

The Lagrangian for the linear sigma model, including only the meson degrees of 
freedom and after the explicit inclusion of the chiral symmetry breaking term, 
can be written as~\cite{Lee}
\be
   {\mathcal{L}}&=&\frac{1}{2}\left[(\partial{\mathbf{\pi}})^2 +
                (\partial\sigma)^2 - m_\pi^2{\mathbf{\pi}}^2 - 
                m_\sigma^2\sigma^2\right]\nonumber\\ 
                &-&\lambda^2 f_\pi\sigma (\sigma^2 + {\mathbf{\pi}}^2) -
                \frac{\lambda^2}{4}(\sigma^2 + {\mathbf{\pi}}^2)^2\, ,
   \label{lagrangian}
\ee
where $\mathbf{\pi}$ and  $\sigma$ are the pion and sigma fields,
respectively, and the coupling $\lambda^2$ is given by
\be
   \lambda^2=\frac{m_\sigma^2-m_\pi^2}{2f_\pi^2}\, .
   \label{coupling}
\ee
From Eq.~(\ref{lagrangian}), we can obtain the Green's functions and
Feynman rules to use in perturbative calculations. In particular, 
the bare pion and sigma propagators 
$\Delta_\pi (P)$, $\Delta_\sigma (Q)$ and the bare one-sigma two-pion 
and four-pion vertices $\Gamma_{12}^{ij}$, $\Gamma_{04}^{ijkl}$ are 
given by (hereafter, capital Roman letters are used to denote four 
momenta)
\be
   i\Delta_\pi(P)\delta^{ij}&=&\frac{i}{P^2-m_\pi^2}\delta^{ij}\nonumber \\
   i\Delta_\sigma (Q)&=&\frac{i}{Q^2-m_\sigma^2}\nonumber \\
   i\Gamma_{12}^{ij}&=&-2i\lambda^2 f_\pi\delta^{ij}\nonumber \\
   i\Gamma_{04}^{ijkl}&=&-2i\lambda^2(\delta^{ij}\delta^{kl}\nonumber\\
   &+&\delta^{ik}\delta^{jl} + \delta^{il}\delta^{jk})\, .
   \label{rules}
\ee
These are sufficient to obtain the modification to the 
pion propagator, both at zero and finite temperature, at any given 
perturbative order.

Alternatively, we can exploit the relations that chiral 
symmetry imposes among different n-point Green's functions. These
relations, known as chiral Ward identities (ChWIs), are a direct
consequence of the fact that the divergence of the axial current may be used
as an interpolating field for the pion~\cite{Lee}. Thus, one could construct
the modification to one of the above Green's functions at a given perturbative
order and from there, build up the induced modification to other Green's
functions related to the former by a ChWI. For example, two of the ChWIs
satisfied --order by order in perturbation theory-- by the functions 
$\Delta_\pi (P)$, $\Delta_\sigma (Q)$, $\Gamma_{12}^{ij}$ and 
$\Gamma_{04}^{ijkl}$ are
\be
   f_\pi\Gamma_{04}^{ijkl}(;0,P_1,P_2,P_3)&=&
   \Gamma_{12}^{kl}(P_1;P_2,P_3)\delta^{ij}\nonumber\\
   &+&\Gamma_{12}^{lj}(P_2;P_3,P_1)\delta^{ik}\nonumber\\
   &+&\Gamma_{12}^{jk}(P_3;P_1,P_2)\delta^{il}\nonumber\\
   \Gamma_{12}^{ij}(Q;0,P)=
   f_\pi^{-1}[\Delta_\sigma^{-1}(Q)&-& 
   \Delta_\pi^{-1}(P)]\,\delta^{ij},\nonumber\\
   \label{Ward}
\ee
where momentum conservation at the vertices is implied, that is
$P_1+P_2+P_3=0$ and $Q+P=0$.

At one loop and after renormalization, we recall that the sigma propagator is 
modified by finite terms. At zero temperature this modification is purely 
imaginary and its physical origin is that a sigma particle, with a mass larger 
than twice the mass of the pion, is unstable and has a (large) non-vanishing 
width coming from its decay channel into two pions. At finite temperature the 
modification results in real and imaginary parts. The real part modifies the 
sigma dispersion curve whereas the imaginary part represents a temperature 
dependent contribution to the sigma width. 

In an expansion in the parameter $m_\pi^2/m_\sigma^2$, the leading
contribution to the sigma self-energy comes from the sigma
polarization bubble where the intermediate particles are pions.
Working in the imaginary-time formalism of Thermal Field Theory,
this is given as $6\lambda^4f_\pi^2I(\omega,q)$
where the function $I$ is defined by 
\be
   I(\omega,q)&=&T\sum_n\int\frac{d^3k}{(2\pi)^3}
   \frac{1}{\omega_n^2 + k^2 +m_\pi^2}\nonumber\\ 
   &&\frac{1}{(\omega_n - \omega )^2 + 
   ({\mathbf{k}} - {\mathbf{q}})^2 + m_\pi^2}\, .
   \label{sigmaself}
\ee 
Here $\omega = 2m\pi T$ and $\omega_n = 2n\pi T$ ($m$, $n$ integers) are
discrete boson frequencies and $q=|{\mathbf{q}}|$. From Eq.~(\ref{sigmaself})
we obtain the time-ordered version $I^t$ of the function $I$, after 
analytical continuation to Minkowski space. The imaginary part of $I^t$ is 
given by
\be
   {\mbox I}{\mbox m}I^t(q_0,q)&=&
   \frac{\varepsilon (q_o)}{2i}\left[
   I(i\omega\rightarrow q_o + i\epsilon ,q)\right.\nonumber\\
   &-& 
   \left.I(i\omega\rightarrow q_o - i\epsilon ,q)\right]
   \nonumber \\
   &=&-\frac{1}{16\pi}
   \Big\{ a(Q^2)\nonumber\\
   &+&
   \left.\frac{2T}{q}\ln\left(\frac{1-e^{-\omega_+(q_0,q) /T}}
   {1-e^{-\omega_-(q_0,q) /T}}\right)\right\}\nonumber\\
   &\times&\Theta(Q^2-4m_\pi^2)\, ,
   \label{imtimeorder}
\ee
where $Q^2=q_0^2-q^2$, $\varepsilon$ and $\Theta$ are the sign and step 
functions, respectively, and the functions $a$ and $\omega_\pm$ are given by
\be
   a(Q^2)&=&\sqrt{1-\frac{4m_\pi^2}{Q^2}}\nonumber \\
   \omega_\pm (q_0,q)&=&\frac{|q_0| \pm a(Q^2)q}{2}\, ,
   \label{imfunc}
\ee
whereas the real part of $I^t$ at $Q=0$ is given by
\be
   {\mbox R}{\mbox e}I^t(0)=-\frac{1}{8\pi^2}\int_0^\infty
   \frac{dk}{E_k}\left[1+2f(E_k)\right]\, ,
   \label{refunc}
\ee
where $E_k=\sqrt{k^2+m_\pi^2}$ and the function $f$ is the Bose-Einstein 
distribution
\be
   f(E_k)=\frac{1}{e^{E_k/T}-1}\, .
   \label{be}
\ee
Therefore, the one-loop effective sigma propagator becomes
\be
   i\Delta_\sigma^\star (Q)&=&\frac{i}{Q^2 - m_\sigma^2 + 
                              6\lambda^4f_\pi^2I^t(Q) }\, .
   \label{newsigprop}
\ee
The temperature-independent infinities are absorbed into the redefinition of 
the physical masses and coupling constants by the introduction of suitable
counterterms in the usual manner.

In order to preserve the ChWIs expressed in Eq.~(\ref{Ward}), the 
corresponding one-loop effective one-sigma two-pion and four-pion vertices
are
\be
   i\Gamma_{12}^{\star\ ij}(Q;P_1,P_2)&=&-2i\lambda^2 f_\pi\delta^{ij}
   \nonumber\\
   &&\left[1 - 3\lambda^2I^t(Q)\right]\, ,\nonumber \\
   i\Gamma_{04}^{\star\ ijkl}(;P_1,P_2,P_3,P_4)&=&-2i\lambda^2
   \times\nonumber\\
   \{[1 - &3&\lambda^2I^t(P_1\!+\!P_2)] \delta^{ij}\delta^{kl}
   \nonumber\\
   +[1 - &3&\lambda^2I^t(P_1\!+\!P_3)]\delta^{ik}\delta^{jl} 
   \nonumber\\
   +[1 - &3&\lambda^2I^t(P_1\!+\!P_4)]\delta^{il}\delta^{jk}\}.
   \nonumber\\
   \label{newvertices}
\ee
The functions in Eq.~(\ref{newvertices}) 
arise from considering all of the possible one-loop contributions 
to the one-sigma two-pion and four-pion vertices, when maintaining
only the zeroth order terms in a systematic expansion in the 
parameter $m_\pi^2/m_\sigma^2$.

\section{Dispersion relation}

We now use the above effective vertices and propagator to construct the
one-loop modification to the pion self-energy. Keeping only the
leading order contribution when considering $m_\pi$, $T$ and $P$ as
small compared to $m_\sigma$ and to zeroth order in
$m_\pi^2/m_\sigma^2$ for which 
\be
   \lambda^2\left(1-\frac{2\lambda^2f_\pi^2}{m_\sigma^2}\right)\approx
   \frac{m_\pi^2}{2f_\pi^2}\, , 
   \label{approx}
\ee
the pion self-energy can be written as
\be
   \Pi(P)&=&\left(\frac{m_\pi^2}{2f_\pi^2}\right)
   T\sum_n\int\frac{d^3k}{(2\pi)^3}\frac{1}{K^2+m_\pi^2}\nonumber\\
   &\Big\{&5 + 2\left(\frac{P^2+K^2}{m_\pi^2}\right)\nonumber\\
   &-&\left(\frac{m_\pi^2}{2f_\pi^2}\right)
   \left[9I^t(0) + 6I^t(P+K)\right]\Big\}\, ,\nonumber\\
   \label{self}
\ee
with $K=(\omega_n,{\mathbf{k}})$ and $P=(\omega,{\mathbf{p}})$. The
pion dispersion relation is thus obtained from the solution to
\be
   P^2 + m_\pi^2 +{\mbox R}{\mbox e}\Pi(P)=0\, ,
   \label{dispersion}
\ee
after the analytical continuation $i\omega\rightarrow p_0 + i\epsilon$. 
The temperature dependent infinities contained in Eq.~(\ref{self}) are
exactly canceled by the contribution from the terms necessary to
introduce at one loop to carry the (vacuum) renormalization 
procedure~\cite{Ayala}.

Let us first look at the dispersion relation at leading order, this
results from
\be
   p_0^2 = p^2 + m_\pi^2\left[1 + \xi\ g(T/m_\pi)\right]\, ,
   \label{dis1b}
\ee
where $\xi=m_\pi^2/4\pi^2f_\pi^2\ll 1$ and the function $g$ 
is given by
\be
   g(T/m_\pi)=\int_0^\infty
   \frac{dxx^2}{\sqrt{1+x^2}}
   f\left(m_\pi\sqrt{1+x^2}\right)\, .\nonumber\\
   \label{g}
\ee
This coincides with the result obtained in Ref.~\cite{Gasser} by means
of ChPT. 

We now look at the next to leading order terms in Eq.~(\ref{self}). 
The first of these is purely real and represents a constant, second
order shift to the pion mass squared. The remaining second order term
shows a non-trivial dependence on $P$. It involves  
the function $S$ defined by
\be
   S(P)\equiv T\sum_n\int\frac{d^3k}{(2\pi)^3}\frac{I^t(P+K)}{K^2+m_\pi^2}\, .
   \label{s}
\ee
The integration in Eq.~(\ref{s}) can only be performed numerically. 
Including all the terms, the dispersion relation up to next to leading order,
for $T\sim m_\pi$ and in the small momentum region is obtained as the 
solution to
\be
   p_0^2&=&p^2 +
   \Big\{1 + \xi g(T/m_\pi) + 
   \frac{\xi^2}{2}g(T/m_\pi)\nonumber\\
   &\times&[9h(T/m_\pi)-4g(T/m_\pi)]
   \Big\}m_\pi^2\nonumber\\
   &+&\xi^2\tilde{S}(p_0,p)\, .
   \label{disp}
\ee 
where the function $h$ is given by
\be
   h(T/m_\pi)\equiv\int_0^\infty\frac{dk}{E_k}f(E_k)\, .
   \label{h}
\ee
and $\tilde{S}(p_0,p)=-(24\pi^4){\mbox {Re}}S^r(p_0,p)$. Figure~1 shows 
the temperature dependence of $m_\pi$ obtained as the solution 
to Eq.~(\ref{disp}) in the limit when $p$ goes to zero. Notice 
that the second order correction in the parameter $\xi$ has 
the same sign as the first order correction. This result
is opposite to the temperature behavior of the pion mass found in 
Refs.~\cite{{Schenk},{Toublan}}.
\EPSFIGURE[h]{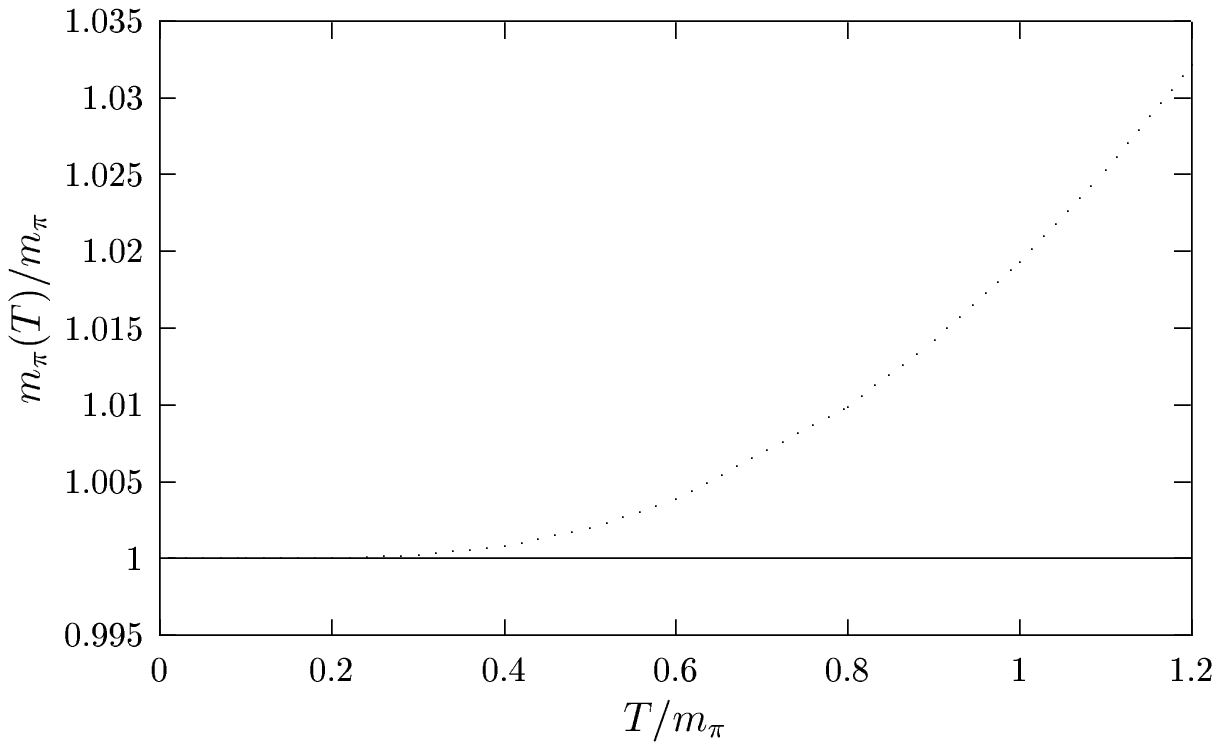,width=2.5in,height=1.75in}{Temperature
dependence of $m_\pi$ to second order in the parameter $\xi$. 
For comparison, the vacuum value is shown as a horizontal line.}

\section{Damping}

Damping is due to two possible kinds of processes: formation of
resonances and pion elastic scattering. At temperatures on the order
of the pion mass, pion elastic scattering is subdominant since
scattering among pions nearly disappears as the relative momentum
approaches zero. In the linear sigma model, this feature is explicitly
realized, for the kinematical regime discussed here, by the
description of pion elastic scattering as a second order process in
the small parameter $\xi$. We start by first looking at resonance
formation. This process is the inverse of the one describing sigma
decay (except that the former does not happen at T=0). 
The relevant diagram to compute
resonance formation is the {\it sunset} pion self energy with an
intermediate sigma line and its explicit expression is
\be
   \Pi_1(P)=4\lambda^4f_\pi^2I(P;m_\pi,m_\sigma)\, ,
   \label{self1}
\ee
where $I$ is the function defined in Eq.~(\ref{sigmaself}) that we
rewrite here in terms of the quantities $E_\pi=\sqrt{k^2+m_\pi^2}$ and
$E_\sigma=\sqrt{({\mathbf{k}}-{\mathbf{p}})^2+m_\sigma^2}$ as
\be
  I(P;m_\pi,m_\sigma)&=&T\sum_n\int\frac{d^3k}{(2\pi)^3}\,\frac{1}
  {\omega_n^2+E_\pi^2}\nonumber\\
  &&\frac{1}{(\omega_n-\omega)^2+E_\sigma^2}\, .
  \label{newI}
\ee
The decay rate $\Gamma^>_1$ is given in terms of ${\mbox I}{\mbox m}\Pi_1$ by
\be
   \Gamma^>_1(p)=-\frac{e^{p_0/T}}{(e^{p_0/T}-1)}\,
   \frac{{\mbox I}{\mbox m}\Pi_1(p_0,p)}{p_0}\, .
   \label{decayr}
\ee
where $p_0=\sqrt{p^2+m_\pi^2}$. ${\mbox I}{\mbox m}\Pi_1$ contains
both creation and decay rates and the factor 
\be
   e^{p_0/T}/(e^{p_0/T}-1)\nonumber
\ee
in Eq.~(\ref{decayr}) eliminates the piece describing pion creation.
We can now compute the mean free path $\lambda$ for a pion traveling 
in the medium before forming a sigma resonance. This is given in terms 
of $\Gamma^>_1$ by
\be
   \lambda = \frac{v}{\Gamma^>_1(p)}
   =-\frac{(e^{p_0/T}-1)}{e^{p_0/T}}\,
   \frac{p}{{\mbox I}{\mbox m}\Pi_1(p_0,p)}\, ,
   \label{freep}
\ee
where $v=p/p_0$ is the magnitude of the pion's velocity. Figure~2 shows plots 
of $\lambda$ for three different temperatures and a value of $m_\sigma = 600$ 
MeV as a function of the pion momentum. $\lambda$ reaches a maximum for
$p\sim 0.4 m_\pi$. Notice that the position of the maximum is approximately
independent of $T$. Except for the location of the maximum, the behavior of
the curves is both in qualitative and quantitative agreement with the 
corresponding result obtained in ChPT~\cite{Gasser}.
\EPSFIGURE[h]{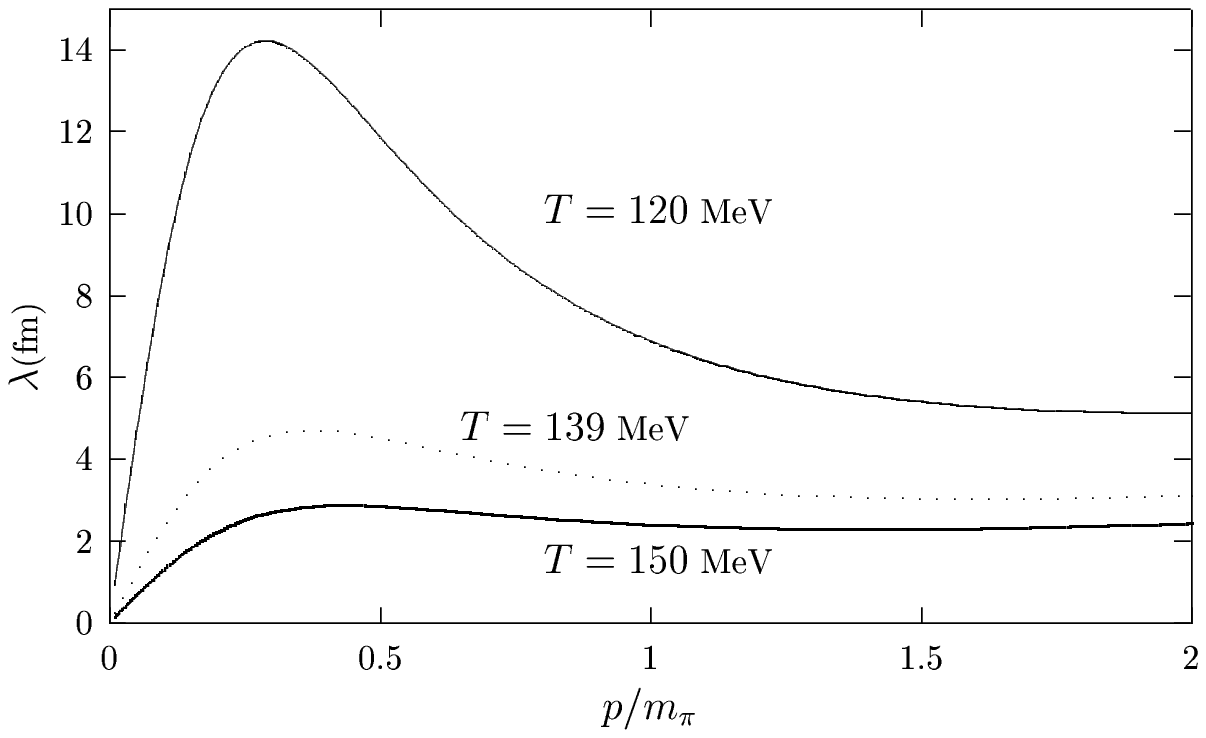,width=2.5in,height=1.75in}{Mean free
path as a function of momentum for a pion to travel the medium before
forming a sigma resonance for $m_\sigma=600$ MeV.}

The pion elastic scattering rate is given in terms of the imaginary
part of the two-loop pion self-energy $\Pi$ given in Eq.~(\ref{self})
\be
   {\mbox I}{\mbox m}\Pi(p_0,p)&=&-24\pi^4\xi^2
   {\mbox I}{\mbox m}{\mathcal S}^r(p_0,p)\nonumber\\
   \Gamma_2(p_0,p)&=&-\frac{1}{p_0}{\mbox {Im}}\Pi(p_0,p)
   \label{imaS}
\ee
and as anticipated, this rate turns out to be about two orders of
magnitude smaller than $\Gamma_1$, given that it is proportional to
$\xi^2$~\cite{Ayala}.

\section{Conclusions}

In conclusion, working in the linear sigma model at finite temperature, we 
have found effective one-loop sigma propagator and one-sigma two-pion and 
four-pion vertices satisfying the ChWIs when maintaining only the
zeroth order terms in an expansion in the 
parameter $m_\pi^2/m_\sigma^2$. We have used these objects to compute the
two-loop order correction to the pion propagator in a pion medium for small 
momentum and for $T\sim m_\pi$ and showed that the linear sigma model yields 
the same result as ChPT at leading order in the parameter 
$\xi=m_\pi^2/4\pi^2f_\pi^2$. We have shown that, contrary to the result in 
Refs.~{\cite{Schenk,Toublan}, the two-loop order correction to the 
pion mass is proportional to $\xi^2$ and that this correction is 
of the same sign as the one-loop correction. The shape of 
the dispersion curve is not significantly altered in the kinematical
regime considered. The results for pion damping are also in
quantitative agreement with ChPT when using a value for $m_\sigma=600$ MeV.

\acknowledgments 
Partial support for this work has been received by CONACyT M\'exico
under grant number 32279-E.

\end{document}